\documentstyle[12pt,aasms4]{article}

\slugcomment{AJ accepted}

\received{15 Dec. 1998}

\accepted{30 Mar. 1999}

\begin{document}

\title{The BHK Color Diagram: a New Tool to Study Young Stellar Populations}
\author{Daniel Devost\altaffilmark{1}}
\affil{D\'epartement de physique and Observatoire du mont M\'egantic,
Universit\'e Laval, Qu\'ebec, QC G1K 7P4, Canada \\
Space Telescope Science Institute, 3700 San Martin Drive, Baltimore, MD  21218
USA}
\authoremail{ddevost@phy.ulaval.ca}

\altaffiltext{1}{Visiting astronomer, Canada-France-Hawaii Telescope, which
is operated by the National Research Council of Canada, the Centre
National de la Recherche Scientifique of France, and the University
of Hawaii.}

\begin{abstract}

A new method to derive age differences between the various super star
clusters observed in starburst galaxies using the two color diagram
(B$-$H) $vs$ (H$-$K) is presented. This method offers a quick and easy
way to differentiate very young and intermediate age stellar
populations even if data on extinction are unavailable. In this case,
discrimination of regions younger and older than 4 Myr is feasible.
With the availability of data on extinction, the time resolution can
be improved significantly.  The application of the method to the
starbursting system Arp 299 is presented.  The validity of the method
is confirmed by comparing the equivalent width of the H$\alpha$ line
with the chronological map of the northern part of NGC 3690.
\end{abstract}

\keywords{Stars: formation --- Infrared: Galaxies: interacting Arp 299
--- Galaxies: individual NGC 3690 --- Galaxies: starburst --- Galaxies:
photometry}

\section{INTRODUCTION}

Prior to the pioneering study of ``flashing galaxies'' by Searle, Sargent , \& 
Bagnuolo (1973), galaxy evolution was thought to be a smooth, exponentially 
decaying process lasting several billion years. The discovery of a significant
population of galaxies actively
forming stars dramatically changed this picture (Harris \& van den
Berg 1981; Seiden \& Gerola 1982; Schweizer 1986; Burstein 1986;
Kumai, Basu , \& Fujimoto 1993). Many galaxies are now known to experience
a series of episodes of enhanced star formation during their lifetime,
especially through merging events (Schweizer 1986; Zepf, Geisler, \&
Ashman 1994), but also following the formation of a bar (Phillips 1996;
Martin \& Friedli 1997). 

Although known to be very important in the evolution of galaxies, the star
formation process is usually poorly accounted for in galactic evolution
models. ``Recipes'' (e.g. Kennicutt 1998) are provided to account for the
global characteristics of the star formation rate versus the interstellar
medium (ISM); those are derived from the global 
or azimuthally averaged properties of observed
galaxies while star formation is a local phenomenon. Little is known about
the local aspects of the star formation process.
The way the star formation propagates, the dynamics of the star formation,
and its relation to the physical properties of the ISM are usually
parameterized as a percolation phenomenon (Mueller \& Arnett 1976;
Gerola \& Seiden 1978; Seiden, Schulman , \& Gerola 1979; Seiden \& Gerola
1982) or regulated by feedback or porosity
(Navarro \& Steimetz 1997; Silk 1997).
All are ad-hoc parameterizations of a very complicated process. 

One of the most difficult tasks in the quantification of the star formation
process is to assign precise ages to star forming regions.
Starburst galaxies are ideal objects to study the local properties of
various star forming regions and their effects on the properties of a galaxy. 
Starbursts form many compact ($\sim$ 2-5 pc), luminous and blue
($-$11 $>$ M$_{UV}$ $>$ $-$18) super star clusters (SSC; Meurer {et~al}. 1995)
that radiate tremendous amounts of light in all wavebands. Also, a
reasonable fraction of this light can easily ionize many atoms
to the first or second stage of ionization, while the high supernovae
rate associated with starbursting regions
excites and ionizes elements to  even higher stages.

These characteristics of starbursting regions provide the opportunity
to study the physical properties of the gas. The knowledge of the gas
properties is essential if one wants to constrain the parameters
involved in the quantification of the star formation process. Several
parameters of a stellar population other than age, can cause a
variation of the integrated magnitude and colors of a SSC. Extinction,
metallicity and the mass of stars formed at present or at any given
time by the cluster are all factors that can affect its observed
characteristics. It is a difficult task to disentangle which parameter
or combinations of parameters cause the differences in the observed
integrated magnitudes and colors of stellar clusters. For example, a
young cluster embedded in dust can be mistaken for an old, unreddened
cluster.

Nevertheless, with appropriate tools, the effect of these various 
parameters can be separated. In this paper, I will show
that the extinction degeneracy can be lifted by a reasonable amount by studying
the SSC integrated light with the (B$-$H) vs. (H$-$K) two color diagram
(BHK diagram).  I discuss in \S 2 the main physical
arguments that support this choice of colors and I propose an
application of the method to the interacting system Arp 299 in \S 3.

\section{CHRONOMETRY WITH THE BHK DIAGRAM}

\subsection{MODELING}

Figure \ref{fig1} shows the theoretical behavior of the BHK colors
for an instantaneous coeval burst
of star formation forming 10$^6$ M$_\odot$ of stars ranging from
1 to 120 M$_\odot$ according to a Salpeter initial mass function
using the models of Leitherer et al. (1999). These models are
implemented with the new set of stellar evolution models of the Geneva
group. For masses of 12~--~25~M$_\odot$ (depending on metallicity)
and above, the Leitherer et al. models are using evolutionary
tracks with enhanced mass loss of Meynet et al. (1994). The tracks
of Schaller et al. (1992), Schaerer et al. (1993a, 1993b), and
Charbonnel et al. (1993) with standard mass loss are used between
12 and 0.8~M$_\odot$. Atmosphere models are from Lejeune, Buser, \& Cuisinier
(1997) for stars which are not in a Wolf-Rayet phase. Wolf-Rayet stars
are modeled with the atmospheres of Schmutz, Leitherer, \& Gruenwald (1992).
A discussion of the uncertainties related to the modeling technique can
be found in Leitherer et al. (1999).

A modification to the models was necessary since stellar evolution
models fail to reproduce the red supergiants (RSG) features at low
metallicities. As the study of Origlia et al. (1998) pointed out, the
low metallicity models does not reproduce the CO 1.62$\mu$ and CO
2.29$\mu$ indices as well as the J$-$K colors of a selected sample of
LMC young clusters.  However, they pointed out that if the fraction of
time spent as a RSG during the core-helium phase is forced to at least
50 \% and if the RSG temperature is maintained to less than 4000 K,
the models agrees well with the observations. Our modeling technique
was modified according to this prescription.

The calculated paths on the BHK diagram of Figure \ref{fig1} are for
metallicities of 0.05 Z$_{\odot}$ and Z$_{\odot}$ and no extinction is
shown.  In the next three sections, I discuss how different
fundamental parameters like the metallicity, extinction, dust emission
and nebular emission can affect the locus of the observed points on
the BHK diagram and why the B, H and K colors are the best choice for
this type of study.

\subsection{THE EFFECT OF METALLICITY}

For a single metallicity, the colors are distributed in a u-shaped
form (see Figure~\ref{fig1}), where two different branches can be
easily seen.  The ``young branch'' runs almost diagonally from the
right to the left of the diagram. The light originating within a young
starbursting region may be dominated by massive O stars ($\sim$80 \%) and 
nebular emission ($\sim$20 \%) in the B band, while in the H and K bands,
the nebular emission may contribute up to as much as $\sim$70 to 85 \% of the total flux.
As the nebular emission becomes negligible with the massive O stars evolving
to become red supergiants, a turnoff occurs in the integrated colors. The color
path then takes the ``old branch'', a path that mainly runs along the
(B$-$H) axis.  After 10 Myr, the color path turns around and follows the
old branch.  This behavior causes a degeneracy between regions aged between
6 and 10 Myr and those aged between 10 and 50 Myr. As we will see in section
\S 3.2, this degeneracy can easily be lifted with spectroscopic data.
After 50 Myr (not shown in Figure \ref{fig1}), the colors do not change
significantly and the evolutionnary path remain in the same region
than the 50 Myr point.

Models in Figure \ref{fig1} also show that an increase in the
metallicity of the SSC produces a different behavior of the colors
(Figure \ref{fig1}). This behavior can be mistaken for an age change.
However, known abundance properties in galaxies
help; the metal distribution in small starburst galaxies
shows no or very shallow abundances gradients
(Kobulnicky \& Skillman 1997; Devost, Roy, \& Drissen 1997; Kobulnicky 1998).
The two metallicity tracks shown in Figure~\ref{fig1} would correspond to
an extreme gradient of several dex kpc$^{-1}$; such a change in
metallicity between two close SSC belonging to the same galaxy is
most unlikely. Thus, the tracks of a group of SSC from the same
galaxy should show parts falling on the same track on the
BHK color diagram, in the absence of extinction.

\subsection{THE EFFECT OF EXTINCTION AND DUST EMISSION}

The extinction vector on the BHK diagram is more or less parallel to
the old branch since E(H-K) $\sim$ 0.1 E(B-H). This alignment is the
key point to extinction correction. Extinction will
move any points in the old or in the young branch along a vector
that is parallel to the old branch. This has two important advantages.
First, the structure of the old branch will still be recognizable since
its alignment with the extinction vector will
prevent its spread by local extinction differences. Global
dereddening is then possible by fitting together the theoretical
old branch path and the observed old branch structure. The second
benefit is that extinction of points belonging to the young branch
will not be moved to the old branch by extinction. An example
of such a degeneracy is given in Figure \ref{fig2}.
This two color diagram plots the behavior of the (B$-$J) vs. (J$-$H)
with the same model parameters as in Figure \ref{fig1} for solar
metallicity stars. We can see that the young and old branches are
still there but extinction of points belonging to the young branch
will cross the path of the old branch, making the distinction
between an age and an extinction effect almost impossible. Also,
the misalignment of the any of the two branches with the extinction
vector causes a spreading of the points by local extinction
differences and it is very unlikely that the structure of
any of the two branches will be conserved.

Warm dust emission will also modify the behavior of the observed infrared
colors of SSCs. Satyapal {et~al}. (1995) showed that
warm and hot dust emission
contaminates only the K band by adding at most 20\% to its
total flux.  Thus, hot dust will make the observed points redder
by a factor of at most 0.2 mag along the
(H$-$K) axis. This value is not high enough to produce a degeneracy
between the young and old branches but will move the points belonging
to the young branch further to the red thus enhancing the difference
between the two branches.

\subsection{THE EFFECT OF NEBULAR LINES}

Nebular emission is also a factor that can produce degeneracies.  In
the infrared bands, the main contribution comes from the Br$\gamma$
emission line which contaminates the K band.  The models of Leitherer
et al. (1999) show that for a continuous burst of star formation, the
equivalent width of the Br$\gamma$ line decreases from $\sim$ 500 \AA\
when the burst is younger than 3 Myr to less than 100 \AA\ when the
burst is older than 4 Myr.  This means that the Br$\gamma$ nebular
line will add $\sim$ 0.12 mag to the K band light when the cluster is
younger than 3 Myr and less than 0.02 mag when the cluster is older
than 4 Myr. The implications for the BHK diagram is that the
Br$\gamma$ emission line pushes the points belonging to the young
branch further to the red and thus, once again, enhancing the
difference between the old and the young branch.

The nebular component in the optical part of the spectrum is much more
important. The hydrogen and the oxygen lines can make an important
contribution to the measured flux and produce unwanted
degeneracies. The H$\alpha$ line contaminates the R band filter while
the H$\beta$ and [O {\sc iii}]($\lambda\lambda$5007,4959) can make
significant contribution to the V band flux. However, the B band is
affected to a lesser degree than the V band filter since the
transmission of the B filter at the wavelength of these lines is about
15\% less than the transmission of the V filter. Also, even moderate
redshifts will shift these lines further into the V waveband. So for
our purpose, the B band filter is minimizing the contamination by the
nebular lines.

\subsection{TIME RESOLUTION}

The considerations outlined in the latter sections restrict the choice
of colors to B, H and K. The alignment of the old branch with the
extinction vector as well as the effects of dust emission and nebular
line emission are strong arguments in favor of this choice.  The
achievable time resolution of this method is mainly limited by the
extinction differences from one cluster to the other. However, even
with no data on extinction, discrimination of regions younger and
older than 4 Myr is still feasible. The age determination in this case
is solely based on the color differences between the old and young
branches.  Nevertheless, if the exciting stars are not completely
embedded in dust and are themselves visible in the optical domain,
nebular lines ratios measured from line-of-sight spectra to the
stellar knots allow a first-order correction. In such a situation,
extinction correction using the Balmer lines decrement helps to lift a
good part of the (B$-$H) degeneracy, then allowing age determinations
with a time resolution limited by the assumed dust and gas properties.

Uncertainties in the dust modeling mainly comes from the assumptions
made on its spatial distribution. Dust extinction can be modeled
assuming a foreground uniform screen or a clumpy layer in front the
extincted stellar population, or assuming that the dust is mixed
within the stellar population (Natta \& Panagia 1984, Calzetti, Kinney
\& Storchi-Bergmann 1994).  Observations of starburst galaxies seem to
rule out dust and stars mixed internally. The intrinsic nature of a
starburst, e.g. the high stellar winds and the frequent supernovae
explosions, may in fact favor the foreground screen geometry (Calzetti
1998). However, very young clusters can be mixed with dust and suffer
very high extinction (Calzetti et al. 1997). This type of regions
maybe missed by the BHK diagram since they will not be detected in the
broad band images.

Nevertheless, for the regions detected in the broad band images, the
worst case scenario would be to make a correction for extinction in
the BHK diagram for a clumpy medium using the foreground screen
approximation. In that case, for a value of A$_V$ $\sim$ 2, the
difference between assuming a foreground screen and a clumpy dust
distribution produces a color difference $\Delta$E(B$-$V) of 0.35. This
produces some uncertainty on the dating of the burst, the magnitude of
which depends on where the burst is along its evolutionary track in
Figure~\ref{fig1}.  A small change in (B-H) produces a larger time
error for a 1 Myr old burst than it does for a 6 Myr old burst.

The uncertainty on the dating of the burst will also be affected by
the model used to correct for extinction. There is a fair amount of
evidence that the extinction derived from the stellar continuum is
systematically less than that derived from the nebular gas
(e.g. H$\alpha$/H$\beta$ ratios). This seems to indicate that the dust
is mostly mixed with the gas, and the stars are mostly seen through
regions of lower optical depth.  However, a higher degree of accuracy
can be reached if the ratio of the extinction derived from the stellar
continuum to the value derived from the nebular gas is known. This
ratio is believed to be between 2 and 3 for starburst galaxies (Meurer
et al. 1995, Calzetti, Kinney \& Storchi-Bergmann 1994). Using the
extinction law of Kinney et al. (1994) to scale between
$\Delta$E(B-V) and $\Delta$(B-H) and a conversion factor between the
continuum extinction and the nebular gas extinction of 2.5, the error
produced by the assumed spatial geometry translates in the BHK diagram
to values of $\sim$ 1.4 Myr along the old branch and $\sim$ 2.0 Myr
along the young branch. Changing the conversion factor between the
continuum extinction and the Balmer extinction does not change
significantly the error estimate on the age determination.

For ages ranging from 10 to 50 Myr, the time resolution has a value
worse than 11 Myr.  The type of resolution derived for these ages
or greater, is incompatible with the purpose of dating star forming
regions to infer the properties of the star formation process. This
means that the BHK diagram can only be useful to trace star formation
to a reasonable resolution for regions that are younger than 10 Myr.

\subsection{A DOUBLE BURST MODEL}

I explored the behavior in the BHK diagram of
a model that forms star clusters in two bursts. 
The super star cluster A in NGC 1569 maybe the result of
such a burst. Gonz\'alez Delgado et al. (1997) found that in this
cluster, red supergiants features as well as Wolf-Rayet features were
present and suggested a double burst scenario to explain the
coexistence of these features within the same cluster. However, the
detection of red supergiants and Wolf-Rayet features within the same
spatial location may also be caused by the alignment of two clusters
along the same line of sight (De Marchi et al. 1997).

I modeled the BHK colors of two identical unresolved bursts occurring 1
Myr (double burst model 1) and 3 Myr (double burst model 2) apart. The
integrated colors of the double burst were derived using:
\begin{center}
$({\rm B-H})_{{\rm DB}}=({\rm B-H})_{\rm O} - 2.5log(\frac{1+{\rm A}}{1+{\rm B}})$
\end{center}
where,
\begin{center}
${\rm A}=10^{0.4({\rm M_{B_O}-M_{B_S}})}$ \quad\mbox{and}\quad ${\rm B}=10^{0.4({\rm M_{H_O}-M_{H_S}})}$
\end{center}
Subscript DB means double burst, subscript O relates to the
original burst and subscript S relates to the second burst.
M$_{{\rm B_O}}$, M$_{{\rm B_S}}$, M$_{{\rm H_O}}$ and M$_{{\rm H_S}}$
are the absolute B and H magnitudes of the original and second burst
respectively. The same algorithm was used for the (H$-$K) color.
The physical parameters of the model are the same as the solar
metallicity burst of Figure~\ref{fig1} and no chemical evolution
occurs between the two bursts.

The results are compared to the single burst model in
Figure~\ref{fig3}. The paths of the single burst and the double burst
model 1 are almost identical but differ significantly between the
single burst and the double burst model 2.  When both bursts are at
young age (t $<$ 4 Myr) the age estimate is not affected. At
intermediate age, a shift occurs in the (B$-$H) color so that a 5 Myr
old single burst can be mistaken for an 6 Myr old double burst (model
2; Figure \ref{fig3}). This means that if a single burst approximation
is made when the studied cluster formed in two bursts separated by 3
Myr, the time estimate can be altered at ages between 6 and 10 Myr by
a value of 1 to 2 Myr.  However, if the two bursts are resolved and
occurred at different spatial locations, each stellar populations will
be given an age by the BHK diagram that will not be affected by the
double burst scenario.

\section{A TEST CASE: Arp 299}

I have tested the BHK diagram method of chronometry by exploring the
SSCs in the interacting system Arp 299. This system includes two
galaxies; NGC 3690 and IC694. The assumed distance to this object is
42 Mpc (Nordgen, et al. 1997; H$_0=$75 km s$^{-1}$ Mpc$^{-1}$).
Several UV bright young SSCs (Meurer et al. 1995) are present as well
as bright infrared sources. With their K band image, Wynn-Williams et
al. (1991) identified four bright infrared sources belonging to the
whole system, three of which are associated with NGC~3690 and one
associated with IC 694. More recently, Satyapal et al. (1997)
established that the light coming from Arp 299 can be entirely
explained by a starburst model. I will show how an age determination
of the various sources with the BHK diagram gives insight into the
star formation history of Arp 299.

\subsection{OBSERVATIONS AND DATA REDUCTION}

Observations of Arp 299 were obtained during three different observing
runs.  First, wide band BVRI images were obtained at the 1.6 meter
telescope of the Observatoire du mont M\'egantic (OMM) in March
1996. The detector was the Thompson THX 1024 $\times$ 1024 CCD with
pixel size of 19 $\mu$ and a plate scale of 0.31\arcsec\
pix$^{-1}$. The JHK band images were taken with the
Canada-France-Hawaii Telescope (CFHT) in January 1997 using the
infrared camera MONICA (Nadeau et al. 1994).  Spectroscopy of Arp 299
was also performed at the CFHT with MOS-ARGUS, a fiber optics system
that allows 2-D spectroscopy with a $\sim$ 12\arcsec $\times$
10\arcsec\ field of view and sampling fiber apertures of
0.4\arcsec. The detector used was the STIS2 2K $\times$ 2K thin CCD
with 21$\mu$ pixels.

\subsubsection{Wide band images}

For optical imaging analysis, bias, flatfield and dark images were
taken during the same night the observations were performed.  The
corresponding correction was made for each image of Arp 299.  Flux
calibration was achieved with Feige 56. Exposure times and other
relevant information about the observing run are listed in Table
\ref{tbl-1}. Bad pixels were removed by interpolation.  Images in one
filter were combined using a combination algorithm to eliminate the
signature of the cosmic rays and to increase the signal-to-noise (S/N)
ratio. The mean FWHM of the point-spread function as measured from
stars in the field of the four combined images is $\sim$ 1\arcsec.
The sky conditions the night of the observations allowed reliable
photometry. The images were then aligned together using field
stars. All the successive image reduction steps were performed with
IRAF.

The second set of observations was done at the CFHT in the J, H, and K
bands with the IR camera MONICA (Nadeau et al. 1994) at the f/8 focus
on January 20 1997 (Table \ref{tbl-1}).  The images were corrected for
flatfield and dark current. The flux calibration was made with the IR
UKIRT standard stars FS2, FS21, and FS25. Bad pixels were corrected
with a mask. In each band, four images were combined using a rejection
routine to raise the S/N ratio and remove the cosmic ray
signature. The images were aligned using point sources common to the
three wavebands.  During the observations, the seeing was 0.7\arcsec.
Once again, the sky was clear and allowed for reliable photometric
calibration. Most of the data reduction steps were performed using
software developed by the Montreal group (D. Nadeau).  Steps not
requiring this software were done with IRAF.

Both set of images were calibrated by deriving the zeropoints of the
magnitude system from the flux of their respective standard
stars. With this type of calibration, there are two main sources of
uncertainty in the measured magnitudes. One comes from the uncertainty
on the value of the zeropoints and the other one comes from photon noise.
However, as we will see in \S 3.2, only the bright points were used
for analysis so the photon noise is negligible. In this case, the
error on the derived magnitudes only comes from the zeropoints
uncertainty which is of the order of 0.1 in B and H and 0.2 in K which
translates to an error of 0.15 in (B$-$H) and 0.22 in (H$-$K) (see the
error bars of Figure \ref{fig5}).

The optical and infrared images were then scaled together using the I
band image of the OMM and the J band image of the CFHT. Sources common
to both images were used as reference points.

\subsubsection{2-D Spectroscopy}

Observations with the 2-D spectrograph MOS-ARGUS were done during the
nights of February 27 to March 1 1998 with the B600 grism and the STIS
II CCD at the f/8 focus of the CFHT. The spectral range is from 3600
to 7000 \AA\ and the dispersion is 2.2~\AA\ pix$^{-1}$ (Table
\ref{tbl-1}). MOS-ARGUS is a fiber optic spectrograph that puts 594
fibers arranged in an hexagon shaped bundle that allows for the
collection of that many spectra per exposure.  The fibers are grouped
in 25 rows that contain a number of fibers increasing from 18 to 30,
giving a maximum field of view of 12\arcsec\ $\times$ 10\arcsec. Each
fiber sampled an aperture of 0.4\arcsec\ which corresponds to a
scale of 80 pc at the assumed distance of 42 Mpc.  Data reduction was
done with IRAF procedures that I specially developed for reduction of
the MOS-ARGUS data. A monochromatic image reconstruction software was
built using the software package IDL.  All spectra were bias and
flatfield corrected. Special care was taken to correct for fiber
transmission and wavelength response.

\subsection{ANALYSIS AND RESULTS}

Figure \ref{fig4} shows the K image of Arp 299 labeled with the
conventional symbols of Wynn-Williams {et~al}. (1991).  The data of
Mazzarella \& Boroson (1993) shows surprisingly uniform oxygen
abundances at about a third solar all over the galaxy.  The
logarithmic extinction differences at H$\beta$\ between the various
regions they studied are of the order of $\Delta c =$ 0.2 $-$ 0.3
which corresponds to a $\Delta$(B$-$H) of 0.6.

Figure~\ref{fig5} compares the models with the observed color of the
0.7\arcsec\ $\times$ 0.7\arcsec\ (150 $\times$ 150 pc) aperture points
in the binned images; I chose a threshold in absolute M$_B$ and M$_K$
magnitude to have only points brighter than $-$10 and $-$14,
respectively. Plotting the observed points instead of the integrated
magnitudes ensures that most of the spatial information is extracted
from the observations since a single resolution element covers 150
pc. Ideally, one would want to be able to resolve each individual
cluster and derive the integrated colors.  This would minimizes the
probability of having more than one burst within one aperture and give
the most reliable spatial information. However, such resolution are
unreachable from the ground for Arp 299.

The parameters of the model in Figure~\ref{fig5} are the same as those
shown in Figure \ref{fig1} with the exception that Figure \ref{fig5}
was made using the stellar evolutionary tracks of metallicity 0.40
solar.  Also, dust emission corresponding to 0.2 mag in K was added to
the original evolutionary track in order to compare the data with
more realistic models. The track with dust emission is then shifted of
a value of 0.2 magnitude to the red along the x axis.  The best global
fit of the theoretical tracks to the data gives an extinction
correction E(B$-$V) = 0.16, which is in agreement with the value
derived by Meurer {et~al}. (1995) and with the value inferred from the
data of Mazzarella \& Boroson (1993).

The ages were determined in the following way. All the points that
fall between the two old branches (see Figure \ref{fig5}) but with
(B$-$H) $<$ 1.2 are considered to be aged between 4 and 7 million
years.  All the points with (B$-$H) $>$ 1.2 are given an age of 7 to
10 Myr. Notice that the points belonging to this part of the diagram
could be aged between 10 and 50 Myr (see \S 2.2) but the nebular
analysis (next paragraph) shows that this is not the case. The value
of (B$-$H) = 1.2 is set by the uncertainty on the color that is caused
by the measured extinction differences since in this case, extinction
is the major source of uncertainty.  All the points that lie outside
the two tracks are labeled to be younger than 4 Myr. These points
belong to the young branch and have been shifted to this part of the
diagram by extinction, dust emission and nebular line emission.  The
derived chronological map is shown on Figure \ref{fig6}.  This map
shows that: {\em i}) objects C, B1 and A are the youngest of the
system; {\em ii}) a region to the south west of object A and one to
the south of object C, are approximately 3 Myr older and {\em iii})
object B2 and a region located to the south west of object A are the
oldest regions of Arp 299.

A further test of the validity and of the limitations of this approach
can be made with the 2-D spectroscopy of MOS-ARGUS.  The equivalent
width of H$\alpha$ (EW(H$\alpha$)) is known to be a very strong
function of age, because it traces very young star forming regions
(Leitherer et al. 1999). However, it cannot be used give precise ages
due to geometrical considerations (Devost, Roy, \& Drissen 1997).
Nevertheless, a young region should show up as a high and peaked
region in EW(H$\alpha$). Figure \ref{fig7} shows a contour map of
EW(H$\alpha$) for NGC 3690, derived from MOS-ARGUS spectroscopic
observations superposed to the wide R band image. The contours
indicate a direct correlation between object C and the highest
EW(H$\alpha$) region located to the north of the image.  This spatial
correlation between the stellar light and the nebular light is very
distinctive of young regions.  The region with high EW(H$\alpha$)
located to the west of B2 is also most interesting. This region has a
high value of EW(H$\alpha$) but has a very low underlying continuum.
Careful examination of the HST UV image of Meurer et al. (1995) shows
that this region doesn't contain any UV bright sources and no sources
are detected in our B or in our K band image.  This lack of continuum
light means that this region is probably very young and still buried
in dust. The continuum light is heavily extincted so this region is
not detected in the broad band images. Such a region is then missed by
the BHK diagram and EW(H$\alpha$) is needed to locate them.

Region B2, also seen in the wide R band image, is an older object
according to EW(H$\alpha$) since it is more extended and has a much
smaller value of EW(H$\alpha$). These results also confirm those obtained
previously with the BHK diagram.  Nevertheless, the BHK diagram
analysis alone cannot assign a precise age to region B2 because the
evolutionary paths of Figure \ref{fig5} are degenerate for ages
between 7 to 10 Myr and 10 to 50 Myr.  The EW(H$\alpha$) is necessary
to lift this degeneracy.  The models of Leitherer et al. 1999 shows that
the {\em integrated} EW(H$\alpha$) has a value of $\sim$ 1 \AA\ at 14 Myr
and rapidly becomes undetectable as the stellar population becomes
older. The value of the {\em surface} EW(H$\alpha$) derived in Figure
\ref{fig7} ($\sim$ 5 \AA) around object B2 is too high for this object to
be in the 10 - 50 Myr age range so this region must be in the 7 -10
Myr age range.

\section{DISCUSSION AND CONCLUSION}

The clear distinction between the two branches seen in the BHK diagram
allows for reliable discrimination between regions younger than 4 Myr
and those older than this age, even if no data on extinction is
available.  The BHK diagram clearly categorizes three (A, B1, C) of
the four infrared sources of Arp 299 as being very young ($<$ 4
Myr). Also, with the help of the EW(H$\alpha$), region B2 is
identified as being between 7 and 10 Myr.  The correlation between the
highest EW(H$\alpha$) region and region C, labeled young from the BHK
diagram analysis, supports the validity and usefulness of this diagram
as a tool for determining the relative ages of young stellar
populations in starbursting regions.  Also, a region located to the
west of object B2 has also a high value of EW(H$\alpha$) and is missed
by the BHK diagram. Clearly, very young regions still buried in dust
are very likely to be missed by the diagram since they are not detected
in broad band images.  

The nebular analysis is found to be a useful complementary tool to the
BHK diagram. In addition to the complementary use of the
EW(H$\alpha$), data on extinction with any of the Balmer lines allows
to acheive a better time resolution. This is essential to derive more
precise star formation properties since extinction is the main
contributor to the uncertainty on the time labeling.  In the case of
Arp 299, the extinction differences inferred from the data of
Mazzarella \& Boroson (1993) limits our resolution to $\sim$ 3 Myr.
Also, their longslit data only sampled a small part of the galaxy and
the extinction from the unsampled regions is unknown.  The use of 2-D
spectroscopy will reduce considerably the uncertainty on
extinction. Corrections using the H$\alpha$/H$\beta$\ line ratio will
lower the time resolution down to the limit imposed by the uncertainty
of the assumed dust model for extinction which I have shown to be of
the order of 1.5 to 2 Myr.

\acknowledgments

I thank Gilles Joncas and Ren\'e Doyon who kindly obtained the infrared
images of Arp 299 at CFHT.  Discussions with Jean-Ren\'e Roy, Carmelle
Robert and Claus Leitherer were most helpful. The comments of D. Calzetti
were appreciated. I would also like to thank the
anonymous referee whose suggestions and comments significantly
helped to improve the paper. This research was supported by the
Natural Sciences and Engineering Research Council of Canada, by FCAR
of the Government of Qu\'ebec and by the STScI DDRF. I am thankful to
the STScI, and in particular to Claus Leitherer, for hospitality
during my stay when part of this work was accomplished.

\clearpage

\pagestyle{empty}

\figcaption{\label{fig1}BHK diagram. The colors were computed with
the models of Leitherer {et~al.} (1999) for the evolution of a coeval burst
of star formation forming 10$^6$ M$_\odot$ of stars distributed along a
Salpeter initial mass function. Models with $\alpha =$ 2.35,
M$_{low} =$ 1 M$_\odot$ and M$_{up} =$ 120 M$_\odot$ and metallicities of
0.05 solar and solar are shown. The tracks are labeled from 0 to
10 Myr with 2 Myr time increments. The 0, 4 and 10 Myr time steps
are shown. The extinction vector is shown on the left of the diagram.}

\figcaption{\label{fig2}Two color diagram diagram but this time with
the (B$-$J) vs. (J$-$H) colors and solar metallicity only. The other
parameters are the same as in Figure~\ref{fig1}.  Notice how the
extinction vector and the old branch have a different alignment.
Also, the difference between the young and old branches is much
smaller than in the BHK diagram.}

\figcaption{\label{fig3} Double burst model. Two double burst models
were explored. A double burst model with two bursts separated by 1
Myr (dashed; Model 1) and 3 Myr (dot dash; Model 2). The physical parameters
of each double burst model are identical to those of the solar metallicity
evolutionary path of Figure \ref{fig1}.}

\figcaption{\label{fig4}CFHT K band image of Arp 299. The bright regions
are identified based on the convention of Wynn-Williams {et~al}. (1991).
North is up and east is left. The total field of view is 1\arcmin
$\times$ 1\arcmin.}

\figcaption{\label{fig5} BHK color diagram of Arp 299. The
solid curves are the theoretical model at 0.40 solar metallicity
without dust emission (left) and considering a 0.2 mag increase
in K magnitude due to dust emission (right). 
The observed points shown are those brighter than M$_B$~=~$-$10 and
M$_K$~=~$-$14. The observations have been corrected for
E(B$-$V)~$=$~0.16. For these colors, the extinction vector is parallel
to the old branch.}

\figcaption{\label{fig6}SSC age distribution in Arp 299. The age
map (greyscale) is shown superposed to the contours
of the B image. In this
picture, region A, B1 and C are the youngest ($<$ 4 Myr) while
region B2 is the oldest (7-10 Myr). Notice that object
B1 does not show up in the B image while it is the brightest in K.
North is up and east is left. The extent of the whole image is
$\sim$ 12 kpc at the assumed distance of 42 Mpc.}

\figcaption{\label{fig7}Contour map of EW(H$\alpha$) superposed on the
wide R band image of NGC 3690. The highest region of EW(H$\alpha$) is
coincident with object C, confirming the results of the BHK diagram
analysis.  Each point on this image is a fiber of 0.4 \arcsec\ for a
total field of view of 12\arcsec $\times$ 10\arcsec. Contours are
labeled with surface EW(H$\alpha$) values ranging from 5 to 53
\AA. North is up and east is left.}

\begin{deluxetable}{cccccc}
\footnotesize
\tablecaption{\label{tbl-1}Observations log}
\tablewidth{0pt}
\tablehead{
\colhead{Band} & \colhead{Number of} & \colhead{Exposure} &
\colhead{Plate} & \colhead{Dispersion} & \colhead{FWHM psf} \nl
\colhead{} & \colhead{exposures} & \colhead{time(s)} &
\colhead{scale(\arcsec/pix)} & \colhead{(\AA/pix)} & \colhead{(\AA)}
}
\startdata
\hline
\multicolumn{6}{c}{OMM imaging} \nl
\hline
B & 3 & 3000 & 0.31 & \nodata & 1.21\arcsec\nl
V & 3 & 1000 &  0.31 & \nodata & 0.97\arcsec \nl
R & 3 & 500  &  0.31 & \nodata & 0.90\arcsec \nl
I & 3 & 500  &  0.31 & \nodata & 1.00\arcsec \nl
\hline
\multicolumn{6}{c}{CFHT IR imaging} \nl
\hline
J & 4 & 1 $\times$ 60 &  0.314 & \nodata & 0.9\arcsec \nl
H & 4 & 6 $\times$ 20 &  0.314 & \nodata & 0.9\arcsec \nl
K & 4 & 9 $\times$ 20 &  0.314 & \nodata & 0.9\arcsec \nl
\hline
\multicolumn{6}{c}{CFHT Spectroscopy} \nl
\hline
3600-7000 \AA & 3 & 3000 & 0.4 & 2.2 & 8 \AA \nl
\enddata
\end{deluxetable}

\setcounter{figure}{0}

\plotone{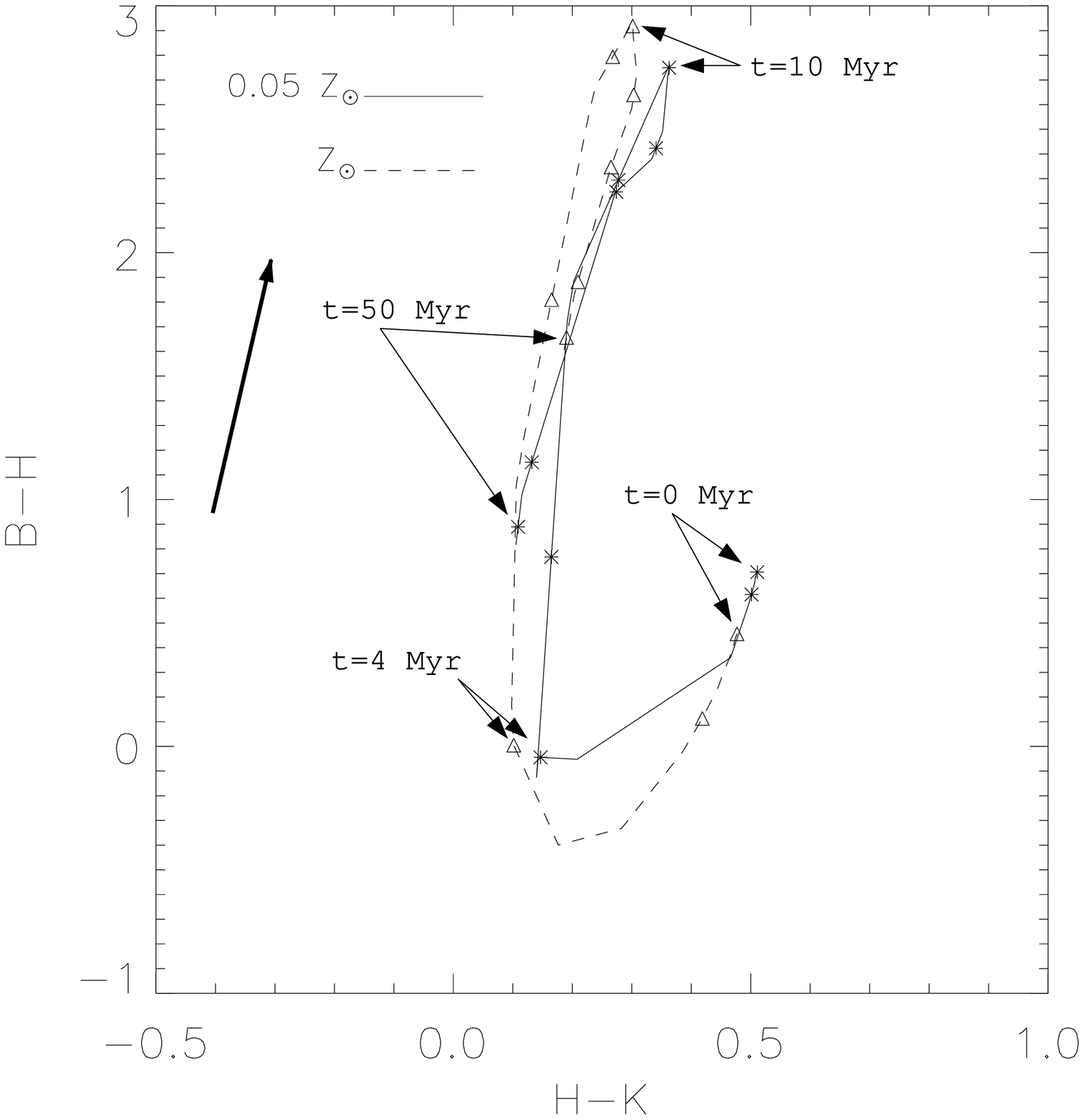}
\figcaption{}

\plotone{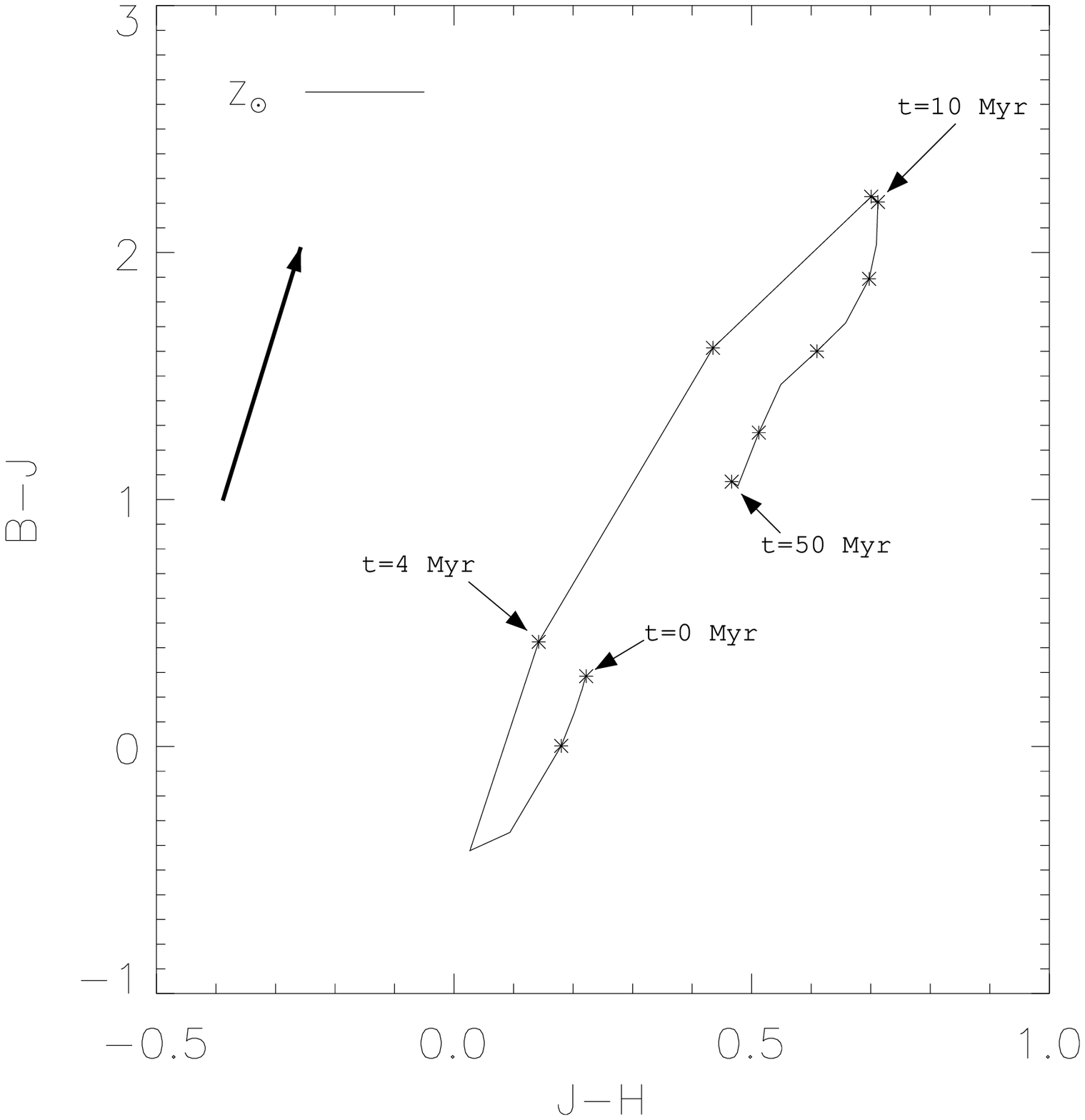}
\figcaption{}

\plotone{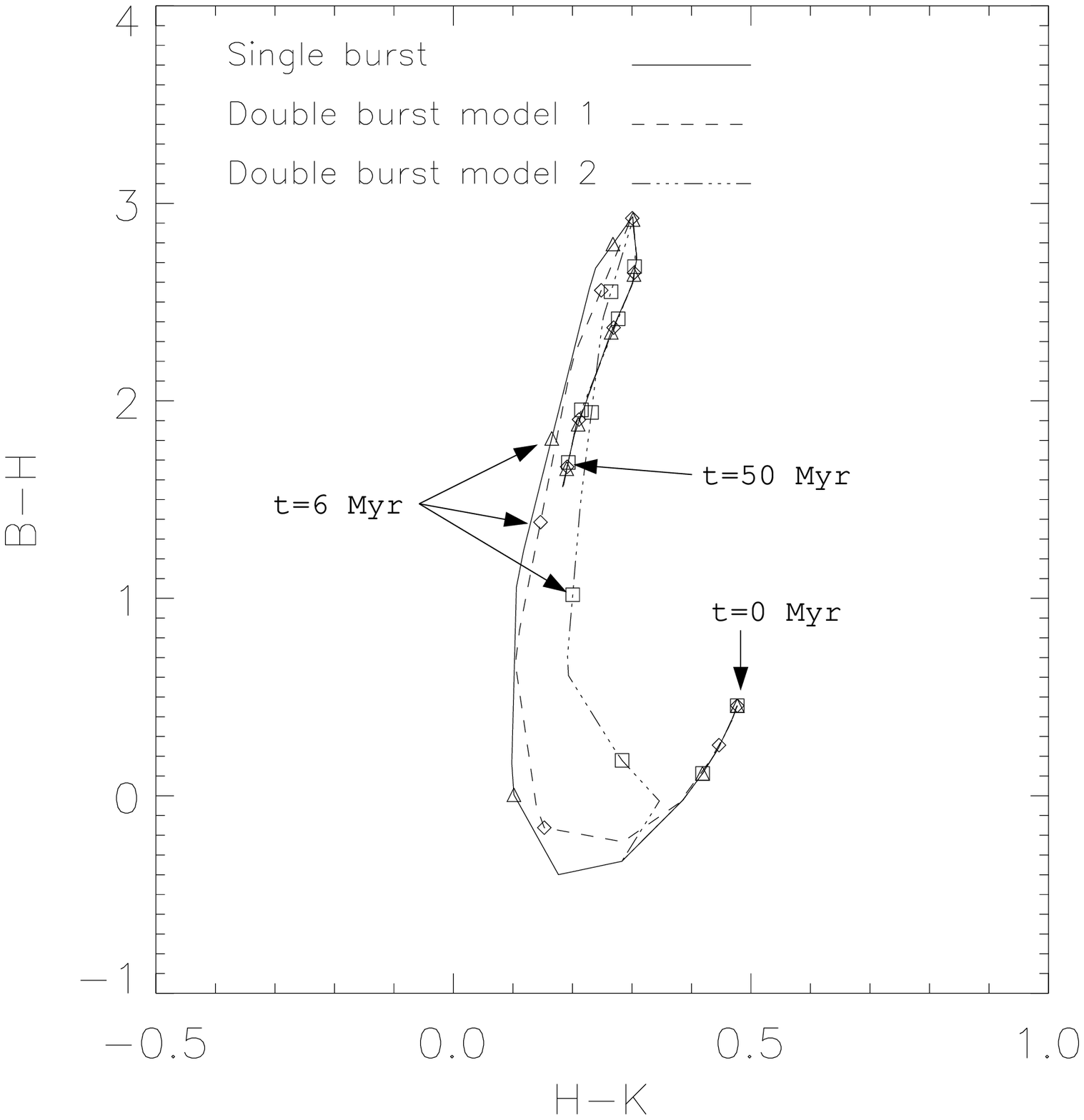}
\figcaption{}

\plotone{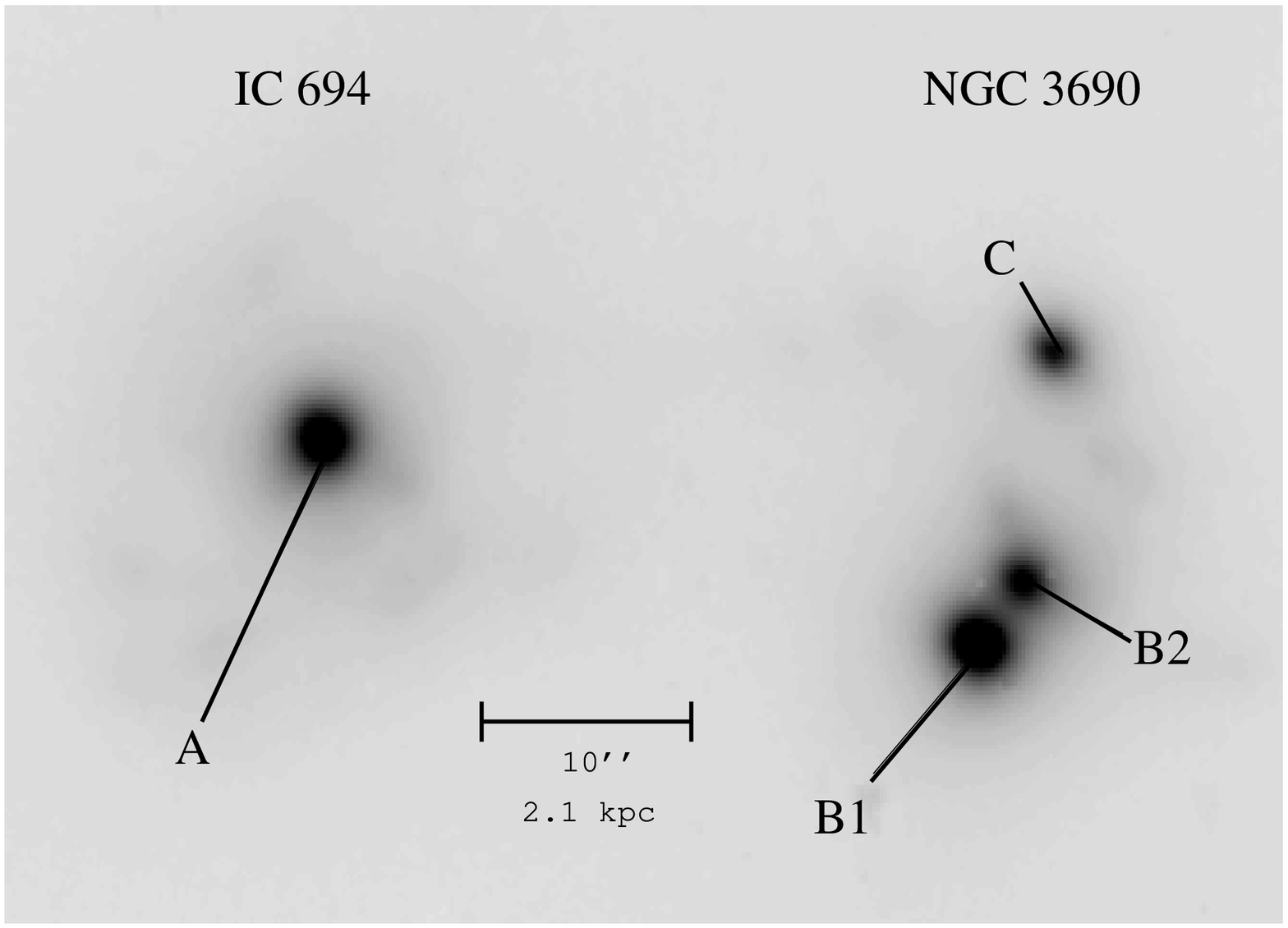}
\figcaption{}

\plotone{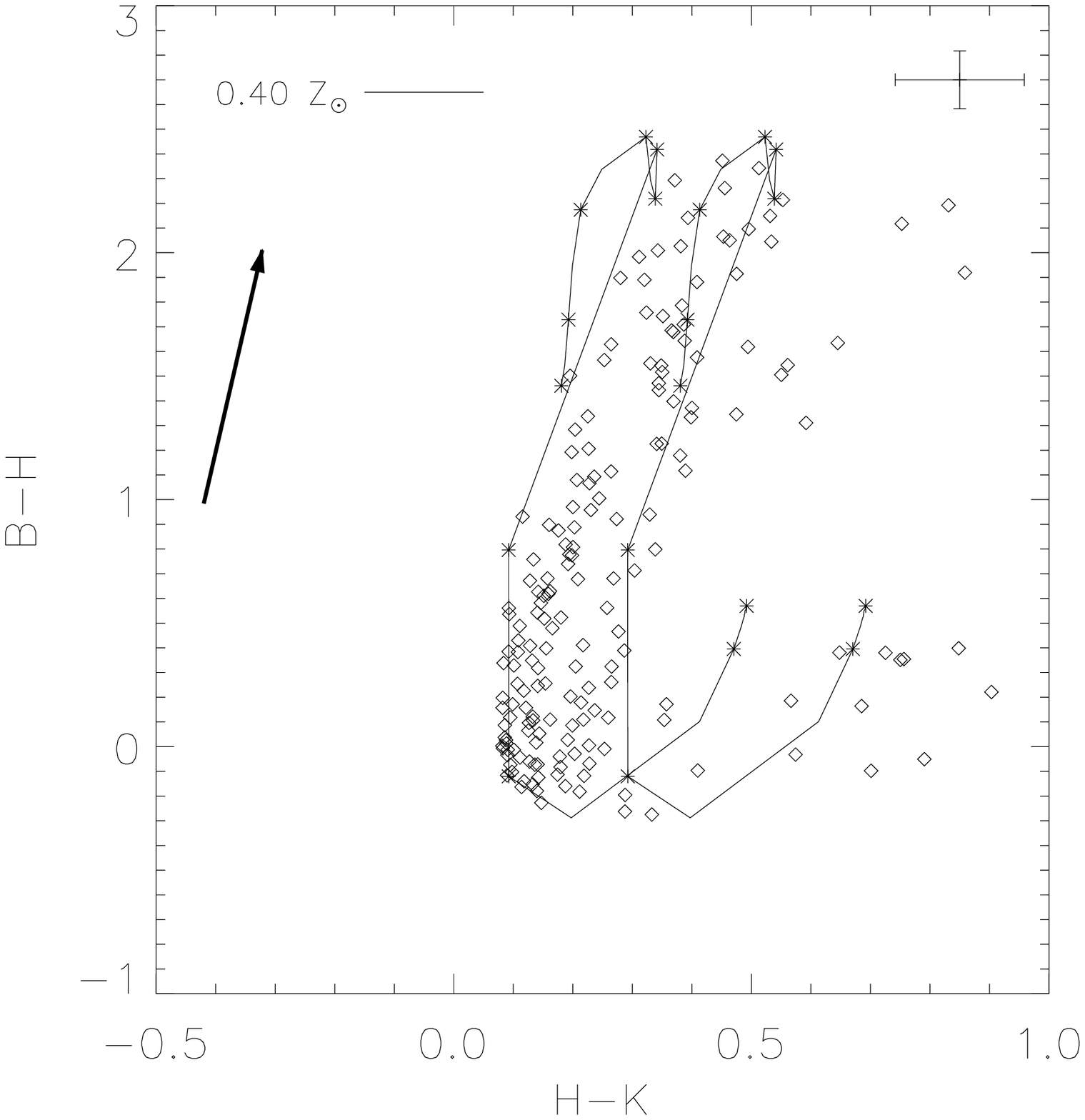}
\figcaption{}

\plotone{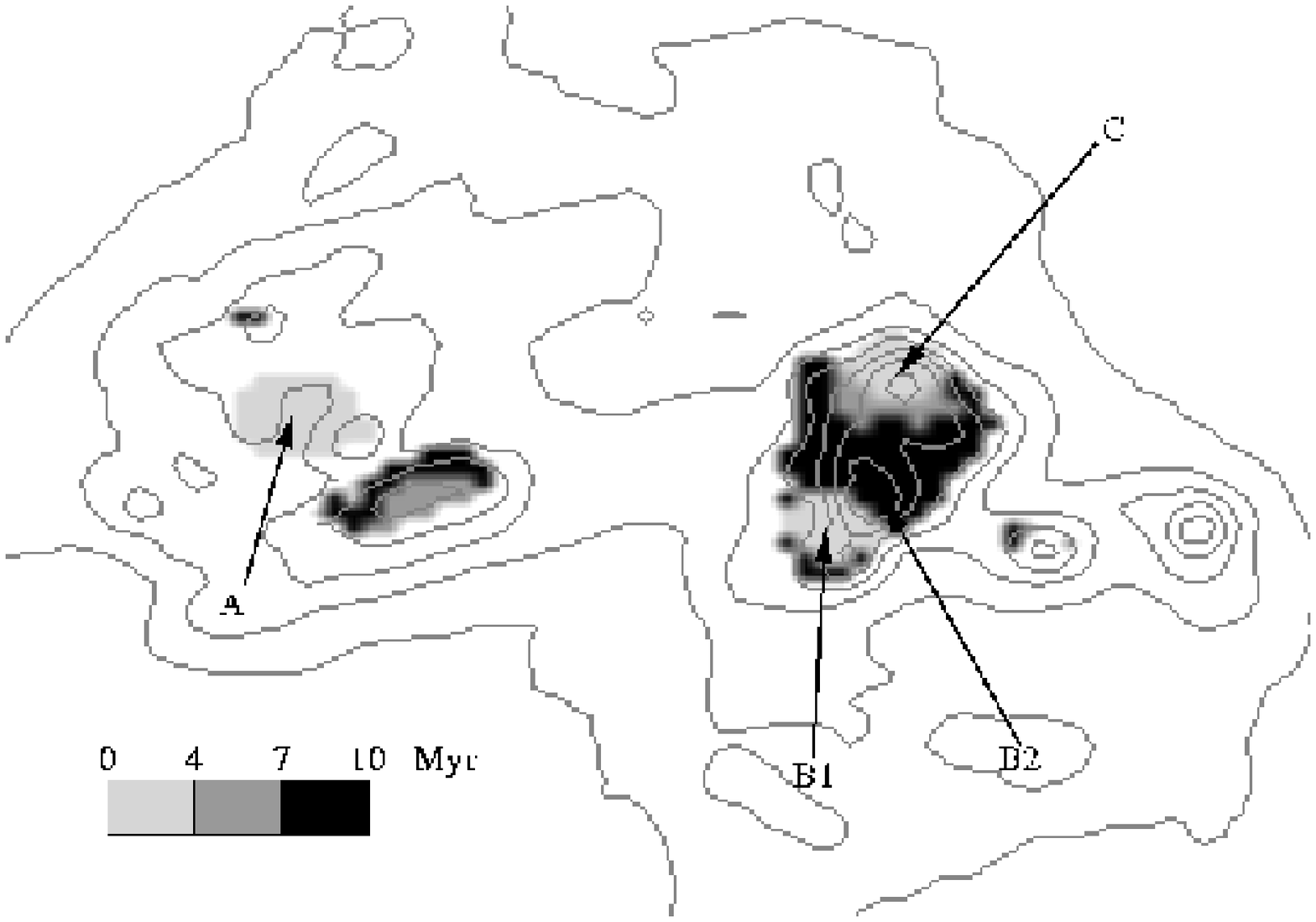}
\figcaption{}

\plotone{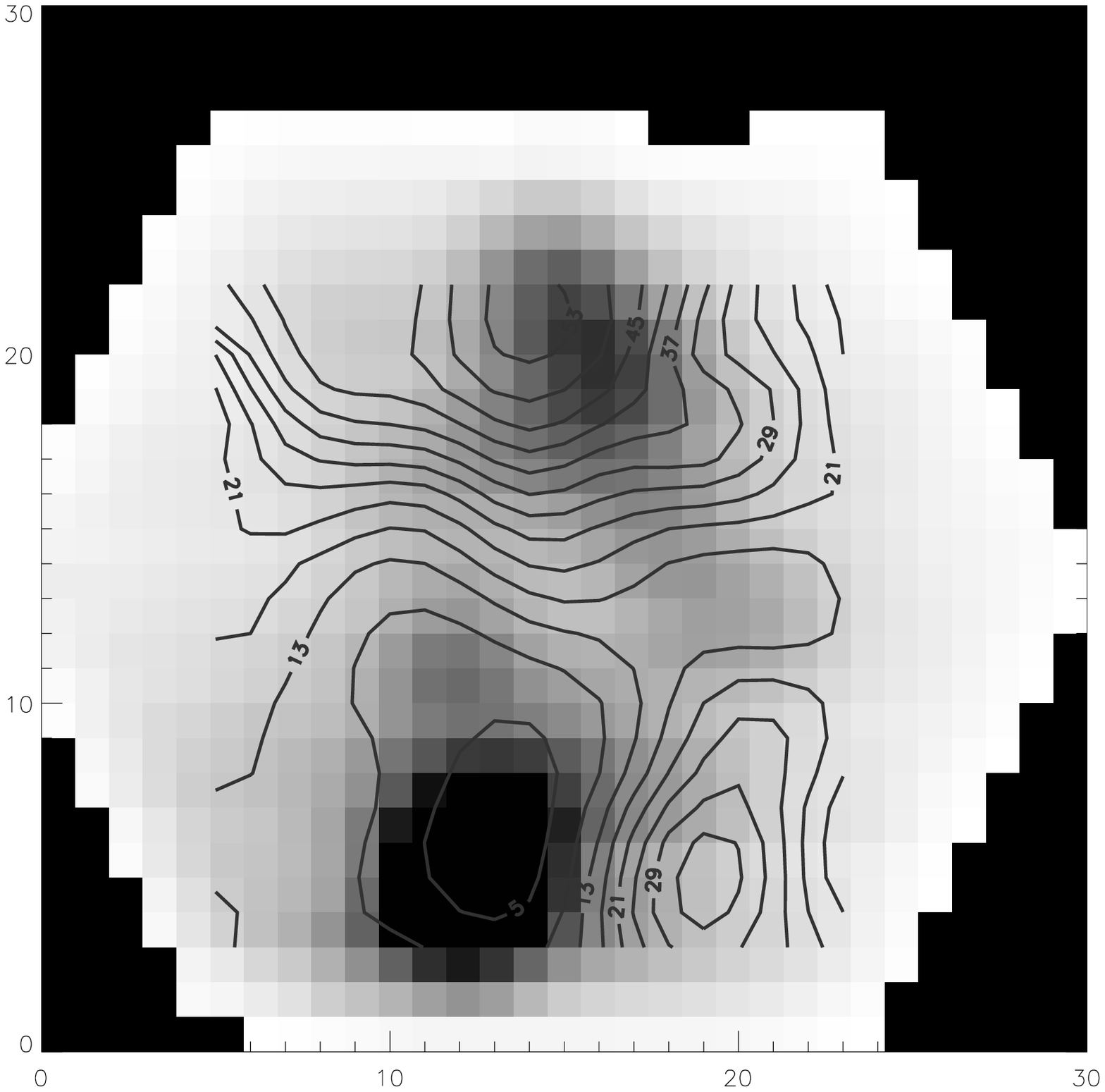}
\figcaption{}

\end{document}